\documentclass[10pt,journal,compsoc]{IEEEtran}

\ifCLASSOPTIONcompsoc
  \usepackage[nocompress]{cite}
\else
  \usepackage{cite}
\fi\ifCLASSINFOpdf
  \usepackage[pdftex]{graphicx}
\else
\fi

\usepackage{amsmath}
\usepackage{comment}
\usepackage{slashbox}
\usepackage{xcolor}
\usepackage{url}

\begin{document}

\title{Comparison of Data Center Cooling Solutions}
\title{The Emergence of Immersion Cooling for (Distributed) Data Centers}
\title{Hot Air Alternatives, the Emergent Role of Immersion Cooling}
\title{A Quantitative Evaluation of Immersion Cooling for Data Centers}
\title{Enough Hot Air:\\ The Role of Immersion Cooling}

\author{Kawsar Haghshenas, Brian Setz, Yannis Bloch, and Marco Aiello,~\IEEEmembership{Senior~Member~IEEE}
\IEEEcompsocitemizethanks{\IEEEcompsocthanksitem Service Computing Department, IAAS, University of Stuttgart, Germany.\protect\\
E-mail: name.lastname@iaas.uni-stuttgart.de}
\thanks{Manuscript submitted May 2022.}}

%
%

\markboth{}%
{Shell \MakeLowercase{\textit{et al.}}: Bare Demo of IEEEtran.cls for Computer Society Journals}
%

\IEEEtitleabstractindextext{%
\begin{abstract}
Air cooling is the traditional solution to chill servers in data centers. However, the continuous increase in global data center energy consumption combined with the increase of the racks’ power dissipation calls for the use of more efficient alternatives. Immersion cooling is one such alternative. In this paper, we quantitatively examine and compare air cooling and immersion cooling solutions. The examined characteristics include power usage efficiency (PUE), computing and power density, cost, and maintenance overheads. A direct comparison shows a reduction of about 50\% in energy consumption and a reduction of about two-thirds of the occupied space, by using immersion cooling. In addition, the higher heat capacity of used liquids in immersion cooling compared to air allows for much higher rack power densities. Moreover, immersion cooling requires less capital and operational expenditures. However, challenging maintenance procedures together with the increased number of IT failures are the main downsides. By selecting immersion cooling, cloud providers must trade-off the decrease in energy and cost and the increase in power density with its higher maintenance and reliability concerns. Finally, we argue that retrofitting an air-cooled data center with immersion cooling will result in high costs and is generally not recommended.

\end{abstract}

}

\maketitle

\IEEEdisplaynontitleabstractindextext
\IEEEpeerreviewmaketitle

\IEEEraisesectionheading{\section{Introduction}\label{sec:introduction}}
\IEEEPARstart{T}{he} scale and the number of data centers are increasing worldwide to meet the rising demand for IT applications and cloud services. High energy consumption (operational cost) and environmental effects are among the main concerns, which both require a great deal of attention. Due to the fact that IT equipment produces heat under load, they have to be artificially cooled to ensure availability and reliability. Therefore, the cooling system is one of the main energy consumers of a data center. Presently, air-cooling is the most prominent technique used in data centers. Using this technique entails that about  40\% of the total energy consumption is dedicated to cooling~\cite{ni2017review}.

In an air-cooled data center, cold air is circulating through the perforated tiles up and into the front of the servers and hot air is pushed out by new cold air coming into the servers. By this method, it is possible to cool server racks with at most 50kW power density~\cite{kheirabadi2016cooling}. In 2018, only 10\% of respondents to a survey reported that the power density of some of their racks was above 40kW~\cite{Smolaks2019}. However, growing power-hungry cluster workloads, such as machine learning ones, produce higher heat densities. These workloads are run on dozens of GPUs, each having thousands of cores that need to be supplied by power. The GPUs increase the power density of the rack, requesting more cooling capacity. Moreover, recent studies show that as the end of Dennard scaling is reached~\cite{dennard1974design}, and transistor sizes approach their practical limitations, new cooling solutions are needed to keep the traditional performance improvement trend. Therefore, CPUs and GPUs with higher power consumption are expected to be manufactured in near future \cite{sun2019summarizing, fan2018analytical, Intel2017}, and consequently, the power density of racks will increase as well. 
 
To address both high energy consumption and low power density problems associated with air cooling, researchers and cloud providers have started to explore alternative solutions. Liquid immersion cooling is a viable one that has attracted attention in the last decade. In immersion cooling, components are fully immersed into a dielectric fluid that conducts heat and does not conduct electricity, therefore, the heat of all IT components is fully removed by liquid, which reduces the Power Usage Efficiency (PUE) of the data center. The PUE is defined as the ratio between total energy and the energy used for the IT equipment of a data center.

The PUE of immersion-cooled data centers is close to perfection, about 1.02-1.04 \cite{matsuoka2017liquid, an20183d, eiland2014flow, chandrasekaran2017effect, shah2018characterizing} which shows that these centers consume 10-50\% less energy compared to their air-cooled counterparts, for the same amount of computational load. In addition, compared to the air, common dielectric fluids have a much higher heat capacity. Therefore, immersion cooling allows for more computing power in less space. While the maximum power density per rack for an air-cooled data center is around 50 KW, immersion-cooled counterpart allows for up to 250 kW per rack \cite{kheirabadi2016cooling, BitFury2015}. 

Immersion cooling is certainly an efficient solution in terms of computing efficiency and power density. In addition, the capital expenditure for constructing a data center, keeping constant the power density, is lower with immersion cooling compared to air cooling~\cite{Bunger2020}. But why it is not deployed in today’s typical data centers? Maintenance and reliability, and specifically the lack of practical information on these aspects, are the primary concerns for adapting immersion cooling technologies~\cite{coles2016immersion, jalilicost, Villa2020, alibaba2018, Varma2019, ramdas2019impact}. Increased number of IT equipment failures, liquid leakage, and liquid evaporation are the main challenges associated with its maintenance procedure, which also impose additional operational costs. However, there are several successful implementations of immersion cooling in data centers. For example, one of the biggest players in cloud computing, Microsoft, has already constructed its first immersion-cooled data center \cite{microsoft2021}.

In this paper, we explore quantitatively the trade-offs between air and immersion cooling technologies and evaluate several aspects of both methods. For this evaluation, we refer to various references, from research studies to practical implementations. Additionally, quantitative analyses on data center efficiency and computing power are presented throughout the paper. This paper will help data center operators to have a better perspective while selecting the best solution for their specific application. Our investigations show that immersion cooling has significant advantages specifically for high power applications and we expect its adoption to grow. Improving the efficiency of a big data center by a small percentage would considerably impact total energy consumption. In addition, the migration from air to liquid cooling could influence operators of smaller data centers in trusting immersion cooling. However, we also argue that retrofitting an air-cooled data center with liquid cooling will result in high costs and is generally not recommended.

The remainder of this paper is organized as follows. The basic concepts of air cooling and immersion cooling, together with a literature review on improving data center efficiency, are presented in Section~\ref{sec:background}.  Then, Section~\ref{sec:practice} contains an overview of several practical implementations of immersion-cooled data centers and several companies that offer immersion cooling server systems. Section~\ref{sec:performance} presents and analysis and comparison of the two cooling solutions including in terms of efficiency, density, cost, and maintenance. Finally,  concluding remarks are provided in Section~\ref{sec:conclusions}.

\section{Background}
\label{sec:background}

Computing hardware consumes electricity, which is dissipated as heat due to resistances in the electrical circuits. In turn, this heat needs to be dispersed to ensure the correct functioning of the equipment. In this section, we provide some background on air cooling and immersion cooling solutions for data centers. In addition, we present a literature review (of reviews) on data centers’ cooling solutions. 

\subsection{Air Cooling}

Data center cooling was not an issue when server rooms were still small and computing hardware did not exceed a rack power density of 5kW~\cite{feelheat20}. These relatively low quantities of heat was removed by basic air conditioning and server fans, which was the easiest solution at the time. Being cheap and easy to implement, air cooling quickly became the standard solution for server cooling and was further developed.

\begin{figure}[htb]
    \centering
    \includegraphics[width=0.45\textwidth]{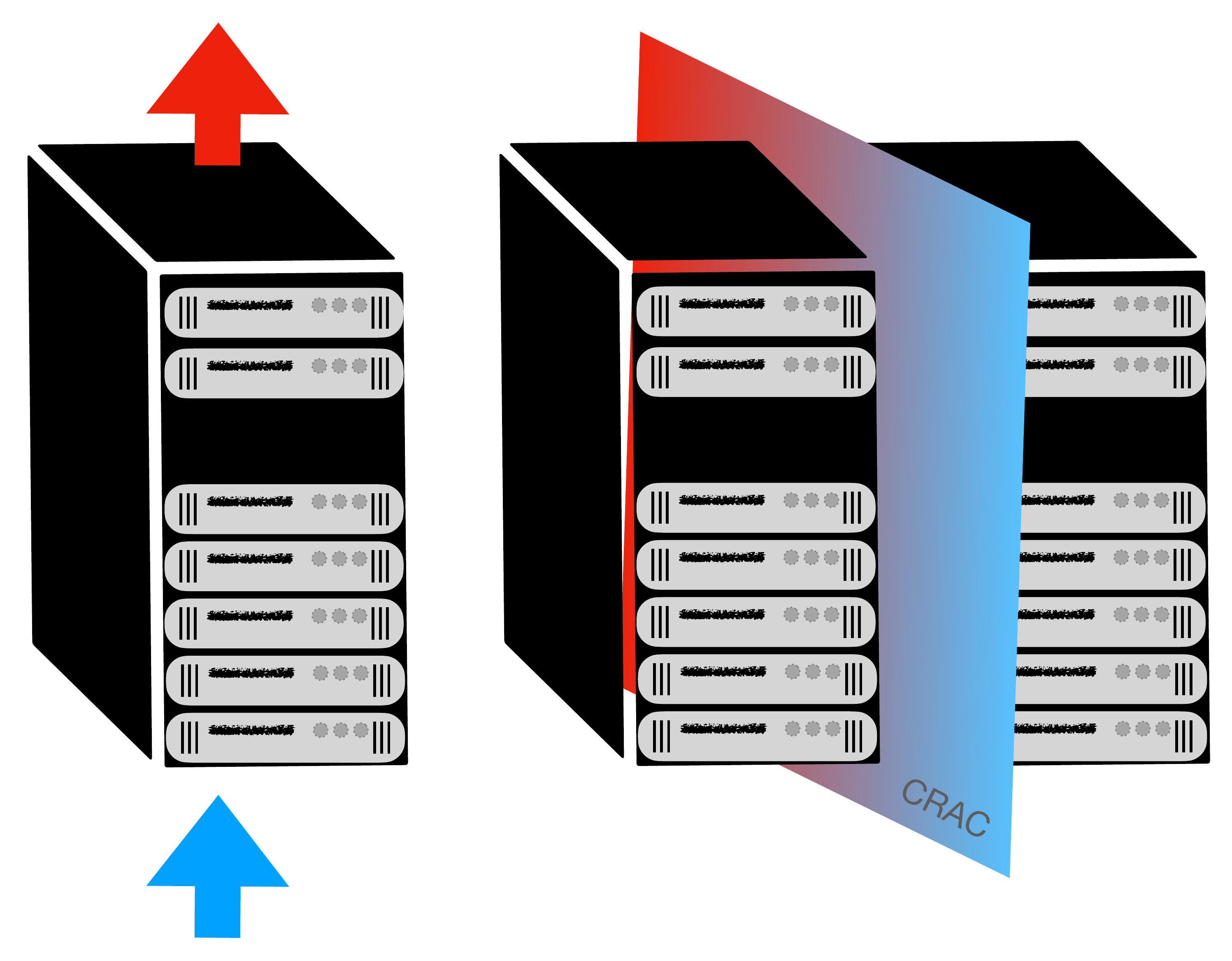}
    \caption{Air-based cooling techniques for data centers}
    \label{fig:sAir}
\end{figure}

It is common for server racks in a data center to stand on a raised floor. A raised floor allows cold air to come from underneath to the front of the servers. Cold air flows through the perforated tiles up and into the front of each server. The fans inside the server chassis suck the cold air inside the servers and then force it over heatsinks covering the processors and other heat-producing hardware. After absorbing the heat of the server components, the warm air is pushed out by new cool air coming into the server.
%
Then, facility fans guide the warm air into a Computer Room Air Conditioning (CRAC) or a Computer Room Air Handling (CRAH), to be cooled back to a low temperature to be used again. The CRAC/CRAH is responsible for cooling the air to a desirable temperature~\cite{geng2014data}. The cooling solutions include chiller-based, water-side economizer, and air-side economizer. A chiller uses electricity and moderately cold water to cool water below the ambient temperature via a compressor. Water-side economizing adds a cooling tower to lower water temperature via evaporation and the air-side economizer uses outside air to reduce temperature. As the next step, the cold air is forced under the raised floor and the cycle restarts from the beginning. In air-cooled data centers, servers need to be set up facing each other for hot/cold aisle separation. This is done to prevent servers from drawing in the exhaust air of other servers. 
Fig.~\ref{fig:sAir} provides a simplified schema of the air cooling approach for racks in data centers. The most basic forms with simple air circulation on the left and the more advanced ones with, e.g., CRAC cooling on the right.

\subsection{Immersion Cooling}

\begin{figure}[htb]
    \centering
    \includegraphics[width=0.40\textwidth]{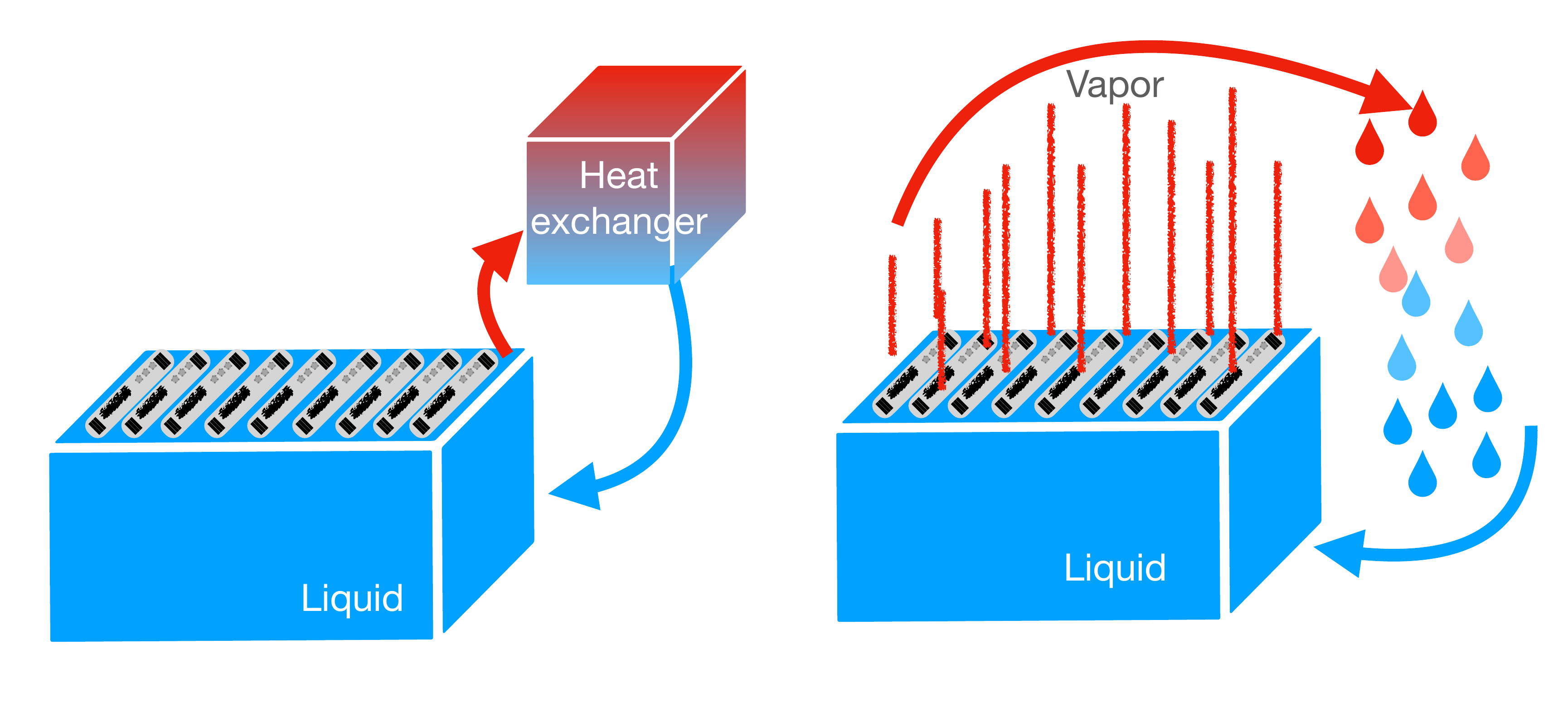}
    \caption{Liquid-based cooling techniques for data centers}
    \label{fig:sLiq}
\end{figure}

Immersion cooling is an approach that uses liquid instead of air to remove heat from computing hardware. In this method, components are fully immersed into a dielectric fluid. A dielectric fluid conducts heat and does not conduct electricity at all; instead, it acts as an insulator. Therefore, the heat of all the components is fully removed by the liquid, eliminating the need for air cooling and moving parts (i.e. server fans). The commonly used dielectric fluids have a much higher heat capacity than air~\cite{shinde2019experimental, eiland2014flow}. 
Most liquids used in immersion cooling are white mineral oil, engineered electric fluids, and other oils~\cite{shinde2019experimental, jalilicost}. There are two types of liquid immersion cooling: single-phase and two-phase, Fig.~\ref{fig:sLiq}.

In single-phase immersion cooling, the fluid stays in liquid form and there is no phase change. Heat emitting components are cooled by the fluid flowing over them, and the heated fluid is transported away. The circulation of the fluid is driven by a pump or by natural convection. In natural convection-driven systems, the heated fluid floats to the top of the tank because it has a higher volume than colder fluid. It then flows, due to more fluid rising to the top, to the side of the tank where it is cooled by a heat exchanger connected to an external loop. The density of the cooled liquid increases again and it sinks back to the bottom of the tank by gravity. In a pump-driven system, the convection is driven by a pump. The pump forces the liquid through an inlet inside the tank and out through an outlet on the opposite side. The liquid is then cooled by flowing through a coolant-to-water heat exchanger. 

In two-phase immersion cooling, the coolant changes phase whenever it gets in contact with a heat-producing component. In order to avoid damaging the components, the boiling point of the coolant has to be lower than the critical temperature of the components. While evaporating on a hot component, the gas will float to the top of the tank and will make room for the new colder liquid coolant. A condenser is located inside the tank above the liquid. Cooling water is flowing through the condenser to transport the heat away. The coolant condenses there and will fall back into the tank where it can absorb heat again~\cite{kanbur2020two}.

\subsection{Literature}

There is a vast body of literature about increasing the efficiency of data centers. These works mostly focus on the scheduling of the workload and its optimization, e.g.,~\cite{DBLP:journals/tsusc/SonDCB17,haghshenas2020infrastructure}. 
The specific aspect of cooling has also attracted some attention and several reviews on cooling solutions for data centers have appeared. One of the first reviews considers the thermal aspects for air cooling in data centers~\cite{patankar2010airflow}. Even though some authors were already advocating a role for immersion cooling~\cite{tuma2010merits}, the review in~\cite{patankar2010airflow} only considers air-based solutions. In~\cite{ebrahimi2014review}, a review of data center cooling technologies is offered, and particular attention is devoted to the opportunity of recovering the heat. For example, district heating can be fed from recovered hot air from a data center.  

Li et al. offer a detailed thermal analysis of cooling solutions, including several strategies based on cold plates, waste heat recovery, and heat pipes~\cite{li2015current}. Immersion cooling is not considered as one of the possible solutions. Kheirabadi and Groulx also focus on the thermal aspects of server cooling~\cite{kheirabadi2016cooling}. They classify the solutions as air-based (CRAC, CRAH, RDHx, and SCHx) and liquid-based (indirect: single-phase, two-phase, heat pipe; and direct: pool boiling, spray cooling, and jet impingement). The results of the review and technology comparison show a higher efficiency for liquid cooling-based solutions while maintaining that air-based solutions are a valid option especially if high efficiency is not a top requirement. The chemical and physical effects of immersion cooling on IT equipment are  investigated in~\cite{shah2016effects}. The results are particularly relevant to understand the maintenance needs and life expectancy of the equipment. In addition, a number of advantages of immersion cooling are identified, such as a decrease in overheating and temperature swings, elimination of fan failures, elimination of dust and moisture-related failures, and reduced corrosion.

A review of the literature and product descriptions is presented in~\cite{WATSON20172711}. The authors select two prototypical systems, one based on air cooling and one on immersion cooling, and provide a comparison of the two based on simulations especially focusing on scalability aspects. The results indicate a preference for immersion cooling. 
A review of thermal management in data centers is offered in~\cite{nadjahi2018review}. The authors recognize that open aisle air-based cooling is the dominant one in practice, though focus on novel techniques and technologies that have good promise in terms of energy efficiency. Kuncoro et al. review exclusively immersion cooling solutions and compare the performance of different types of cooling liquids~\cite{kuncoro2019immersion}. 

Building on the current state of the art, our aim with the present work is to provide a quantitative comparison of cooling solutions from the computational point of view and consider not only issues of energy efficiency, but also computational density, investment, and maintenance costs. To the best of our knowledge, our paper is the first work that compares air and liquid cooling for data centers quantitatively and comprehensively.
\section{Immersion Cooling in Practice}
\label{sec:practice}

Although air cooling is the dominant solution in data centers, several large companies are adopting the immersion cooling technology and a number of startups have appeared offering innovative systems. At the time of writing, \emph{Microsoft} has announced its first liquid immersion-cooled data center in Washington, USA \cite{microsoft2021}. This data center is used for cloud-based communication platforms such as Microsoft Teams.  \emph{Alibaba} also uses single-phase immersion cooling (1PIC) tanks in its data centers. They have shown that using immersion cooling reduces the total power consumption by 36\% and helps to achieve a PUE of 1.07~\cite{alibaba2019}. Another example comes from the \emph{BitFury} group that built a 40+ MW data center that comprises 160 tanks running with a PUE of 1.02 using two-phase immersion cooling (2PIC)~\cite{BitFury2015}.


Furthermore, some companies are offering immersion-cooled server systems. \emph{Asperitas}, for example, is a Dutch company located in Amsterdam that offers complete liquid immersion cooling solutions to its customers. Their server enclosures operate with single-phase immersion cooling and natural convection, thus avoiding the use of any mechanical parts. The immersion-cooled servers of \emph{Asperitas} are insulated; this is done to capture all heat produced by the servers in the fluid and to allow for maximum waste heat reutilization. Each of their AIC24 enclosures can contain up to 48 servers or 288 GPUs with a footprint of only 60cm x 120cm~\cite{Asperitas2021}.  

An alternative interesting approach was that proposed by the Dutch company \emph{Nerdalize}. The idea was to offer a distributed data center by displacing servers in the residential buildings~\cite{ngoko2018future}. The immersion cooled servers would exchange heat with water that was then used for indoor heating and hot tab water. When the energy savings of (water) heating are taken into account, the PUE of such a system would be less than 1.0. The company deployed several servers before bankruptcy in 2018. Interestingly, the company which restarted \emph{Nerdalize}, \emph{LeafCloud}, decided to opt for air cooling, mostly due to its lower maintenance costs and failure rates. 

Another company offering liquid immersion cooling enclosures in various sizes is \emph{Submer} \cite{submer2021}. All of \emph{Submer}’s products use single-phase immersion cooling and are ranging from cabinet-sized enclosures called microPod, up to megaPod, which are set up inside shipping containers. The microPods are capable of cooling 5kW of components even in direct sunlight which makes them suitable for companies who want to cool their in-house equipment efficiently. On the other hand, megaPods are targeted for higher computing powers. They can be put in almost any place since there is only electricity and network connection needed. For example, it would be possible to install a megaPod onto or near a building which then supplies the building with heat. Similar to \emph{Nerdalize} and \emph{LeafCloud}, their products are excellent for waste heat reutilization.


\section{Efficiency, Density, and Cost}
\label{sec:performance}

Cloud providers are generally profit-oriented. This means that data centers are implemented and optimized for maximum computing output using the least amount of investment and running costs. While some operators make decisions with sustainability in mind, the majority of them choose the most cost-efficient solutions. 

In this section, we compare air cooling and immersion cooling solutions on several dimensions including computing efficiency, computing density, power density, cost, and maintenance. From the perspective of a data center operator, there is a clear link between each of these factors and the profit margin. The efficiency of the cooling method affects each of the mentioned factors, in turn reinforcing the importance of choosing the appropriate cooling solution.

\subsection{Computing Efficiency}

\begin{table}[!t]
\caption{PUE values reported in the literature}
\label{tab:pue}
\renewcommand{\arraystretch}{1.3}
\centering
\begin{tabular}{l | c | c}
\hline
\textbf{Source} & \textbf{\begin{tabular}[c]{@{}l@{}}Reported PUE\\ \small{air cooling}\end{tabular}} & \textbf{\begin{tabular}[c]{@{}l@{}}Reported PUE\\ \small{immersion cooling}\end{tabular}} 

\\ \hline
 \noalign{\hrule height 1pt}
\textbf{\cite{matsuoka2017liquid}} & 1.1 & \textless{}1.04 


\\ \hline
\textbf{\cite{an20183d}} & & 1.02 
\\ \hline
\textbf{\cite{eiland2014flow}} & & 1.03 - 1.17 

\\ \hline
\textbf{\cite{chandrasekaran2017effect}} & & 1.02 - 1.03 

\\ \hline
\textbf{\cite{shah2018characterizing}} & 1.7 - 2.9 & 1.02 - 1.03 

\\ \hline
\textbf{\cite{Miller2014}} & 1.12, 1.18 &  
\\ \hline
\textbf{\cite{McNevin2013}} & 2.2 - 2.61 & 
\\ \hline
\end{tabular}
\end{table}

\begin{figure*}[!t]
    \centering
    \includegraphics[width=0.85\textwidth]{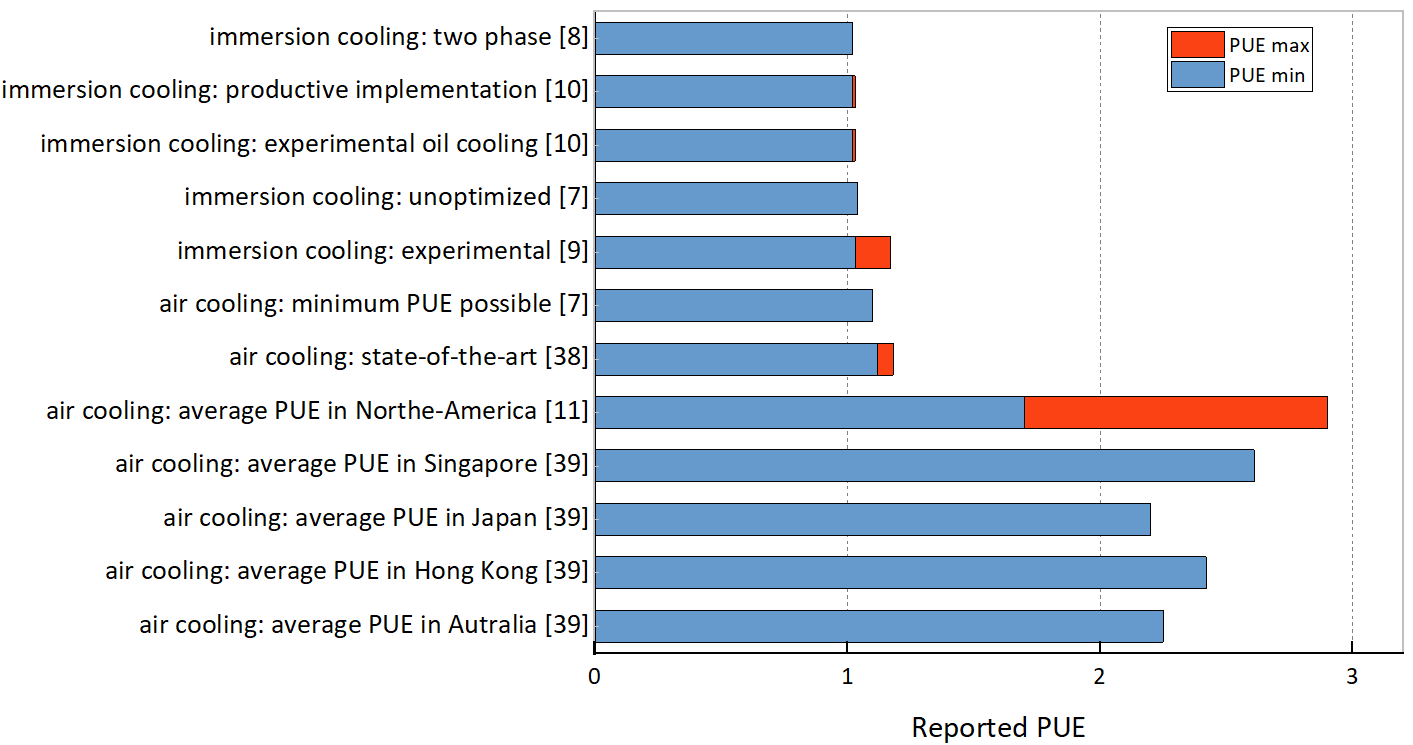}
    \caption{PUE values based on the cooling technology}
    \label{fig:pue}
\end{figure*}

The most popular metric for measuring the energy efficiency of a data center is the PUE~\cite{reddy2017metrics}. PUE, originally proposed in 2006, and standardized as ISO/IEC 30134-2:2016 in 2016. The definition of PUE is given in Equation~\ref{pue}.

\begin{equation}\label{pue}
    \text{PUE} = \frac{\text{total energy}}{\text{IT energy}} \geq 1
\end{equation} 

Since the IT energy consumption is included in the total energy, the value of PUE will typically be greater or equal to one. Several studies report on the PUE of data centers with specific installations of immersion and air cooling solutions, these are shown in Table~\ref{tab:pue}. 

For air-cooled data centers, reported PUE values range from 1.1 to 2.9~\cite{matsuoka2017liquid, shah2018characterizing, McNevin2013, Miller2014}. Values close to 1.1 can only be achieved by hyper-scale data center facilities, which are especially optimized for efficient cooling. For example, the state-of-the-art air-cooled data centers of Google have the PUE of 1.12~\cite{Miller2014}. This means that in these data centers, 89\% of the total energy is consumed by IT equipment. However, high efficient data centers with state-of-the-art design make up for only a small portion of data centers worldwide~\cite{jones2018information}. In addition, average air-cooled data centers have a much higher PUE compared to the most efficient ones. The average PUE has been reported to be between 2.2 and 2.61 for the data centers in Singapore, Japan, Hong Kong, and Australia. This  shows a significant difference between state-of-the-art hyper-scale and average data centers~\cite{McNevin2013}.

Studies regarding immersion cooling data centers have reported consistently better results. For these, the PUE falls in the range  1.02 to 1.04~\cite{matsuoka2017liquid, an20183d, eiland2014flow, chandrasekaran2017effect, shah2018characterizing}. The PUE of 1.02 seems to be the sweet spot for immersion cooling. Table~\ref{tab:pue} shows how the various studies agree on the fact that PUE values around 1.02 are achievable with immersion cooling. With a PUE of 1.02, about 98\% of the energy consumed by the data center goes to the IT equipment. This is close to the perfect efficiency. The maximum reported PUE for immersion-cooled data centers is 1.17~\cite{eiland2014flow}. This case relates to an experiment aiming at maximum cooling capacity without efficiency in mind. For further illustration, Fig.~\ref{fig:pue} show a plot of the values from Table~\ref{tab:pue} with respect to the type of technology used for cooling. 

While PUE is a useful metric for evaluating the efficiency of a data center over time, it is not equally suitable for comparing data centers with one another. Furthermore, it does not give direct information about the computing efficiency of the data center. In~\cite{reddy2017metrics}, we reviewed metrics for data centers and explored refinements and alternatives to PUE. In first approximation, PUE just offers a reasonable indicator of data center efficiency. From a computing perspective, it is also desirable to have a value representing the computing power in relation to energy consumption. Computing performance is traditionally measured by running a set of operations (a benchmark) on a device and measuring the completion time. An example of such a performance benchmark is measuring the number of floating point operations per second.

To compare the efficiency of different systems or facilities, one measures the computing performance of running a benchmark in the average/peak power of the system. This results in a metric of floating-point operations per second per watt (FLOPS/W), a metric capable of showing the performance in relation to the energy consumption. 

In a data center, in addition to the servers for running the operations, we have non-computational equipment such as the cooling system which consumes energy. Therefore, we consider an alternative metric for computing efficiency as follows.

\begin{equation}\label{etta}
    \eta = \frac {\text{computing performance}}{\text{power}} (FLOPS/W)
\end{equation}

The computing performance and the power of IT equipment are relevant for the calculation of FLOPS/W; however, they do not change the overall efficiency while changing the cooling solution. This is because changing the cooling system does not influence the computing performance and the IT equipment's power consumption. Typically, when modifying the cooling systems, the IT energy remains constant while the total energy changes. To calculate the improvement in FLOPS/W by changing the cooling method, the knowledge of PUE and FLOPS/W can be combined for an improved computing efficiency metric, presented as Equation~\ref{etta+}.
\begin{equation}\label{etta+}
    \eta_{\text{data center}} = \frac {\frac{b_0}{b_1}}{\text{IT power}}.\frac{1}{\text{PUE}}
\end{equation}
where $b_0$ and $b_1$ stand for the number of benchmark operations and benchmark time, respectively. The inverse of PUE represents the part of the energy that has been used by IT equipment. For example, a data center with a PUE of 1.5 uses two-third of its energy for IT equipment. By changing the cooling method, the value of $b_0$, $b_1$, and IT power remain the same. This means that overall computing efficiency depends directly on the fraction of power used for IT equipment. Furthermore, percentage differences in FLOPS/W can be calculated having only the PUE.

Knowing this, the expected improvement in the data center's overall computing efficiency by switching from air cooling to immersion cooling can be calculated by comparing 1/PUE values. For example, when migrating from best practice air cooling (PUE=1.12) to immersion cooling (PUE=1.02), the overall computing efficiency is increased by 9.8\%. While this is a significant increase that reduces the operational expenditures, migrating from standard air cooling (PUE=2) to immersion cooling (PUE=1.02) increases the computing efficiency by 96.1\%. Therefore, such a conversion would reduce the energy consumption by half. The increase and decrease in computing efficiency when changing cooling method is shown in Table~\ref{efficiency_improvement}.

\begin{table}[!t]
\caption{Efficiency improvement by switching cooling technique}
\label{efficiency_improvement}
\renewcommand{\arraystretch}{1.3}
\centering
\begin{tabular}{l|c|c|c}
\hline
\backslashbox{From}{To} & \textbf{\begin{tabular}[c]{@{}l@{}}Standard \\ air cooling\end{tabular}} & \textbf{\begin{tabular}[c]{@{}l@{}}State of the art \\ air cooling\end{tabular}} & \textbf{\begin{tabular}[c]{@{}l@{}}Immersion\\cooling\end{tabular}} 

\\ \hline
 \noalign{\hrule height 1pt}
\textbf{Standard air cooling} & -  & 78.6\% & 96.1\%            
\\ \hline
\textbf{\begin{tabular}[c]{@{}l@{}}State of the art\\ air cooling\end{tabular}} & -44\% & - & 9.8\%                    
\\ \hline
\textbf{Immersion cooling} & -49\% & -8.9\% & -                      
\\ \hline
\end{tabular}
\end{table}

Let us present a numerical example based on one of the most powerful server CPUs currently on the market, the AMD EPYC 7742. This processor runs with 225-watt maximum power consumption and achieves around 3.48 TeraFLOPS~\cite{Trader2019}. Therefore, the processor can calculate 15.5 GigaFLOPS per watt at its peak performance:
\begin{equation}\label{etta_EPYC}
    \eta_{\text{EPYC7742}} = \frac {\text{3.48 TFLOPS}}{\text{225W}} = 15.5 \text{ (GFLOPS/W)}
\end{equation}

The central processing unit that calculates the FLOPS, is not the only power consuming component of a server. Therefore, to calculate the server computing efficiency, one needs to know the proportion of power consumed by the processor in relation to the server's total power. The computing efficiency of a server is calculated by:

\begin{equation}\label{etta_server}
    \eta_{\text{server}} = \eta_{\text{processor}}.p_{\text{processor}}
\end{equation}
where $p_{\text{processor}}$ stands for the proportion of power consumed by the processor in relation to the total server consumption. Various works have reported different power breakdowns for the components of a server. In accordance with the results presented in~\cite{gill2018taxonomy}, we assume that 50\% of the server’s power is consumed by its processor. From Equation~\ref{etta_server}, the computing efficiency of the server amounts to 7.73 GFLOPS/W.

To calculate the computing efficiency of the IT equipment, we  consider the power of all IT equipment including storage and network facility. In this way, the computing efficiency of IT equipment is calculated as:
\begin{equation}\label{etta_IT}
    \eta_{\text{IT equipment}} = \eta_{\text{server}}.p_{\text{server}}
\end{equation}
where $p_{\text{server}}$ stands for the proportion of the server's power in relation to the total IT power. Indeed, we use $p_{\text{server}}$ in equation~\ref{etta_IT}, to take into account the energy consumption of non-computational IT equipment (storage and network) in calculating the computing efficiency of IT equipment. The contribution of the servers, storage, and network facility in the total power consumption of a data center has been reported in several works~\cite{dayarathna2015data, shehabi2016united}. For our numerical example, we assume the average of reported values, that is 77\%. Therefore, $\eta_{\text{IT equipment}}$ is 5.95 GFLOPS/W.

Finally, the computing efficiency of the data center is calculated by:
\begin{equation}\label{etta_DC}
    \eta_{\text{data center}} = \eta_{\text{IT equipment}}.\frac{1}{\text{PUE}}
\end{equation}

By Equation~\ref{etta_DC}, the computing efficiency of a data center cooled by standard air cooling, state-of-the-art air cooling, and immersion cooling is calculated as 2.98, 5.32, and 5.84, respectively. 
The computing efficiency of an immersion-cooled data center is almost 9\% higher than a state-of-the-art air-cooled hyperscale data center. For the average data center, the switch to immersion cooling offers even more improvement. Operators can decrease the power consumption by about 50\% without any decrease in computing performance.

\subsection{Computing and Power Density}

\begin{table}[!t]
\caption{Power densities reported in the literature}
\label{tab:density}
\renewcommand{\arraystretch}{1.3}
\centering
\begin{tabular}{l | c | c | c}
\hline
\textbf{Source} & \textbf{\begin{tabular}[c]{@{}l@{}}Air cooling\\ \small{density }\end{tabular}}&\textbf{\begin{tabular}[c]{@{}l@{}}Immersion cooling\\ \small{density}\end{tabular}}& \textbf{\begin{tabular}[c]{@{}l@{}}Normalized\\to kW/l\end{tabular}} 

\\ \hline
 \noalign{\hrule height 1pt}
\textbf{\cite{matsuoka2017liquid}} & &14 kW/bathtub & 0.045 

\\ \hline
\textbf{\cite{gess2014investigation}} & & 400 W/bathtub & 0.23 
\\ \hline
\textbf{\cite{an20183d}} & & 250 kW/rack & 0.14               

\\ \hline
\textbf{\cite{tuma2010merits}} & & 4 kW/l & 4 

\\ \hline
\textbf{\cite{kheirabadi2016cooling}} & 33-50 kW/rack & & 0.018-0.028

\\ \hline
\textbf{\cite{Smolaks2019}} & 40 kW/rack & & 0.022

\\ \hline
\end{tabular}
\end{table}

Computing density is defined as the amount of computation that a system can offer in relation to its size. The metric used to determine the computing density of a data center is $FLOPS/m^3$. This metric considers everything: from the servers to the power management system and up to the cooling equipment. Comparing air-cooled and immersion-cooled in these terms also shows significant differences. The required space for immersion cooling has been reported to be about one third of traditional air cooling method~\cite{matsuoka2017liquid}.

The increase in computing density by immersion cooling comes from various factors. In immersion cooling, there is no space needed in between the servers and racks for airflow. Tubs can stand right next to each other, and the only limitation is the accessibility of the service crew. In addition, there is significantly less cooling equipment needed (only piping), and no raised floors or air vents are required~\cite{matsuoka2017liquid}. 
According to the literature, with immersion cooling the density increases at both the rack and facility levels~\cite{tuma2010merits}. Furthermore, many air-cooled data centers trade their computing density for efficiency and reliability~\cite{Miller2014}. The cooling units of these centers are often bigger than they need to be.

Power density is another important factor for data center operators. Power density must be low in air-cooled data centers, otherwise, either fan speeds need to be increased or air temperature needs to be lowered, where both solutions lower the energy efficiency. According to the research, rack-level power density of air-cooled data centres is about 0.018 kW/l - 0.028 kW/l~\cite{kheirabadi2016cooling}, while the density of immersion-cooled data centres is between 0.045 kW/l~\cite{matsuoka2017liquid} and 0.23 kW/l~\cite{gess2014investigation}. Even more, 4 kW/l has been reported to be possible in immersion-cooled data centers with enough coolant flow and a compact IT design~\cite{tuma2010merits, tuma2010open}. Table~\ref{tab:density} presents the data gathered from various references on the power density of air-cooled and immersion-cooled data centers. Since power densities are presented in kW per rack for air-cooled data centers and for immersion cooling there is no typical rack size, the values need to be converted to kW/l. This metric is the most meaningful one when the whole facility is considered.

As the reported numbers show, the power density with immersion cooling is almost six times higher than air cooling. Therefore, power density can be another strong motivation for data center operators to consider immersion cooling for their future projects, specifically for running their high-performance workloads. Going forward, computing is driving devices into higher power consumption and CPUs and GPUs capable of consuming more power are expected to be manufactured. Given this trend, air cooling can not provide sufficient cooling capacity, while maintaining high power densities.

\subsection{Cost}

An important factor to consider when adopting immersion cooling is capital and operational expenditures. The capital expenditures include the sealed chassis for immersing the IT equipment, dielectric fluids, as well as pumps and tubing. Similar to air-cooled data centers, the operational expenditures include electricity, staff, network connection fees, as well as supporting and maintaining the IT equipment.

According to Bunger et al.\cite{Bunger2020}, for air-cooled data centers with a power density of 10 kW per rack, the capital expenditures are \$7.02 per watt. For a liquid-cooled data center with a similar power density, the cost is reduced slightly to \$6.98 per watt. The benefit of liquid cooling is also that much higher power densities can be achieved. Assuming a power density of 40 kW per rack, the expenditures are further reduced to \$6.02 per Watt. Furthermore, when it comes to the operational expenditures, a reduction between 9-20\% is expected with regards to the energy cost due to the absence of fans. This is in agreement with the results of Neudorfer et al. who state that a 5-10\% reduction of the IT energy consumption is expected due to the avoidance of internal fans~\cite{Neudorfer2016}. The need for fans is also absent for the entire facility, reducing the total energy costs by an amount in the range of 15-25\%. Another point to consider is that the dielectric fluids present in the sealed chassis can be used virtually indefinitely, assuming some form of filtration is present. 

Day et al. highlight that when building a new data center and optimizing it for liquid cooling from the ground up, capital expenditure savings can be achieved over air-cooled data centers~\cite{Day2019}. On the contrary, retrofitting an air-cooled data center with liquid cooling can result in higher costs. An important exception is retrofitting an air-cooled data center with limited floor space and power capacity. In this case, the increased power density possible with liquid cooling, as well as a reduction in energy consumption, can address both the space and power limitations with one solution.

\subsection{System Maintenance}

The operational costs of a data center include system maintenance. The cooling method influences the components’ environmental conditions and consequently affects the number of and time to failures and overall equipment lifetime. 

The number of maintenance requests caused by failures has been reported to be almost 6.6\% higher for an immersion-cooled data center compared to a traditional air-cooled counterpart~\cite{coles2016immersion}. The higher number of maintenance requests associated with immersion cooling impose additional operating costs. In addition, higher failure rates can degrade the components’ lifetime, but immersion cooling can compensate for this degradation with lower junction temperatures~\cite{jalilicost}. Besides the number of maintenance requests, the maintenance procedure is more challenging with immersion cooling. In this case, immersed IT equipment is removed by opening the lid and lifting the equipment out of the tank. This can lead to liquid evaporation.

According to the work presented in~\cite{Villa2020} and~\cite{alibaba2018}, the maintenance overheads and reliability concerns, as well as the leaks and spills, are the top contributors to the low adoption of immersion cooling. Indeed, the enclosure that the racks are immersed in must be sealed perfectly to avoid liquid evaporation/losses. Complete enclosure sealing would mitigate the problem but is not practical. The high number of maintenance requests, compared to the air cooling system, cause access issues and adds operating costs related to compensating the fluid losses~\cite{Varma2019}.

The operating costs imposed by the liquid loss have been evaluated in~\cite{coles2016immersion}. The cost of the lost liquid divided by the cost of the IT equipment’s energy usage has been reported to be 4.68 for a specific implementation. This number shows that the maintenance overhead is a significant drawback associated with immersion cooling. It can even mitigate the energy efficiency improvement of immersion cooling. However, it should be noted that the implementation characteristics including the liquid price and the electricity costs affect the reported value. In~\cite{coles2016immersion}, these values are set at 75 $\$/liter$ and 0.09 $\$/kWh$, respectively. One-phase immersion cooling is reported to have fewer maintenance needs than two-phase immersion cooling~\cite{Varma2019}.

\section{Conclusions}
\label{sec:conclusions}

Air cooling is the traditional solution for addressing the heat dissipation problem of data centers. High energy consumption and low cooling capacity---and consequently limited power density---are the main challenges associated with air cooling. Immersion cooling is emerging as a novel method with many advantages in terms of efficiency, density, and cost. In the present work, we provide a quantitative comparison of these two approaches and overview the results presented in the literature regarding the two alternatives. While most data centers around the world are relying on air cooling, immersion cooling is recognized as a potential alternative and several cloud providers have already constructed their immersion-cooled data centers. The key findings of the present work are listed next.

First, based on the PUEs reported in the literature, we conclude that the immersion cooling method consumes less energy and has a higher computing efficiency. Our analysis shows that a typical immersion-cooled data center consumes almost 50\% less energy compared to its air-cooled counterpart. Second, immersion-cooled data centers allow for much more compact designs, more than three times the density of their air-cooled counterparts. In liquid immersion-cooled data centers, there is no trade-off between efficiency and density. Conversely, air-cooled data centers can only be either-or. Immersion-cooled data centers can be placed in ordinary spaces having lower requirements and the need for additional equipment. Third, while research on immersion cooling is mostly targeting efficiency, the aspect of power density should not be overlooked. The increase of computing power in a specific volume is even more important than the efficiency improvement. The most conservative figures for immersion cooling are about double density in kW/l on rack-level compared to the maximum possible in air-cooled data centers. Resources estimating 4kW/l illustrate what is potentially possible with IT equipment optimized for liquid immersion cooling. This high-density capability makes immersion cooling the first-choice solution for running high-performance workloads. In addition, it allows manufacturing the CPUs and GPUs with higher frequency and higher power consumption.  
Fourth, the capital expenditure for an air-cooled data center with a power density of 10 kW per rack is about 4\% higher compared to its immersion-cooled counterpart. In addition, a 9-20\% operational expenditures reduction is expected with immersion cooling. 
Fifth and final, the main downside of immersion cooling is the challenges related to the maintenance and reliability concerns. This is a reasonable explanation for why even with significant advantages in efficiency and density, not many companies have switched their cooling solution to immersion cooling, yet.

In general, immersion cooling appears to be a solution that should be strongly considered when designing a new data center and we expect its adoption to grow. At the same time, retrofitting any air-cooled data center with liquid cooling does not seem convenient in most cases, if feasible at all. Considering the reliability and maintenance challenges and uncertainties, immersion cooling might not be useful for small and average power densities. On the contrary, immersion cooling appears to be the best solution for high power densities.

\ifCLASSOPTIONcompsoc
  \section*{Acknowledgments}
\else
  \section*{Acknowledgment}
\fi

The presented research is funded by the Netherlands Organisation for Scientific Research (NWO) in the framework of the Indo-Dutch Science Industry Collaboration programme with project NextGenSmart DC (629.002.102).

\ifCLASSOPTIONcaptionsoff
  \newpage
\fi

\bibliographystyle{unsrt} 
\bibliography{biblio,dc}

\begin{thebibliography}{10}

\bibitem{ni2017review}
Jiacheng Ni and Xuelian Bai.
\newblock A review of air conditioning energy performance in data centers.
\newblock {\em Renewable and sustainable energy reviews}, 67:625--640, 2017.

\bibitem{kheirabadi2016cooling}
Ali~C Kheirabadi and Dominic Groulx.
\newblock Cooling of server electronics: A design review of existing
  technology.
\newblock {\em Applied Thermal Engineering}, 105:622--638, 2016.

\bibitem{Smolaks2019}
Max Smolaks.
\newblock Power density – the real benchmark of a data centre.
\newblock
  \url{https://virtusdatacentres.com/item/389-power-density-the-real-benchmark-of-a-data-centre},
  2019.

\bibitem{dennard1974design}
Robert~H Dennard, Fritz~H Gaensslen, Hwa-Nien Yu, V~Leo Rideout, Ernest
  Bassous, and Andre~R LeBlanc.
\newblock Design of ion-implanted mosfet's with very small physical dimensions.
\newblock {\em IEEE Journal of Solid-State Circuits}, 9(5):256--268, 1974.

\bibitem{sun2019summarizing}
Yifan Sun, Nicolas~Bohm Agostini, Shi Dong, and David Kaeli.
\newblock Summarizing cpu and gpu design trends with product data.
\newblock {\em arXiv preprint arXiv:1911.11313}, 2019.

\bibitem{fan2018analytical}
Yuehong Fan, Casey Winkel, Devdatta Kulkarni, and Wenbin Tian.
\newblock Analytical design methodology for liquid based cooling solution for
  high tdp cpus.
\newblock In {\em 2018 17th IEEE Intersociety Conference on Thermal and
  Thermomechanical Phenomena in Electronic Systems (ITherm)}, pages 582--586.
  IEEE, 2018.

\bibitem{Intel2017}
Intel.
\newblock New 2nd generation intel xeon salable processor.
\newblock \url{https://www.intel.com/content/dam/www/public/us/en/
  documents/guides/2nd-gen-xeon-sp-transition-guide-final.pdf}, 2017.

\bibitem{matsuoka2017liquid}
Morito Matsuoka, Kazuhiro Matsuda, and Hideo Kubo.
\newblock Liquid immersion cooling technology with natural convection in data
  center.
\newblock In {\em 2017 IEEE 6th international conference on Cloud networking
  (CloudNet)}, pages 1--7. IEEE, 2017.

\bibitem{an20183d}
Xudong An, Manish Arora, Wei Huang, William~C Brantley, and Joseph~L
  Greathouse.
\newblock 3d numerical analysis of two-phase immersion cooling for electronic
  components.
\newblock In {\em 2018 17th IEEE intersociety conference on thermal and
  thermomechanical phenomena in electronic systems (ITherm)}, pages 609--614.
  IEEE, 2018.

\bibitem{eiland2014flow}
Richard Eiland, John Fernandes, Marianna Vallejo, Dereje Agonafer, and
  Veerendra Mulay.
\newblock Flow rate and inlet temperature considerations for direct immersion
  of a single server in mineral oil.
\newblock In {\em Fourteenth Intersociety Conference on Thermal and
  Thermomechanical Phenomena in Electronic Systems (ITherm)}, pages 706--714.
  IEEE, 2014.

\bibitem{chandrasekaran2017effect}
Sriram Chandrasekaran, Joshua Gess, and Sushil Bhavnani.
\newblock Effect of subcooling, flow rate and surface characteristics on flow
  boiling performance of high performance liquid cooled immersion server model.
\newblock In {\em 2017 16th IEEE Intersociety Conference on Thermal and
  Thermomechanical Phenomena in Electronic Systems (ITherm)}, pages 905--912.
  IEEE, 2017.

\bibitem{shah2018characterizing}
Jimil~Manojbhai Shah.
\newblock {\em Characterizing contamination to expand ASHRAE envelope in
  airside economization and thermal and reliability in immersion cooling of
  data centers}.
\newblock PhD thesis, The University of Texas at Arlington, 2018.

\bibitem{BitFury2015}
Two-phase immersion cooling a revolution in data center efficiency.
\newblock 3M{\texttrademark} Novec{\texttrademark} Engineered Fluids, 2015.

\bibitem{Bunger2020}
R.~{Bunger}, W.~{Torell}, and V.~{Avelar}.
\newblock Capital cost analysis is of immersive liquid-cooled vs. air-cooled
  large data centers.
\newblock {\em Schneider Electric White Paper 282}, 2020.

\bibitem{coles2016immersion}
Henry Coles and Magnus Herrlin.
\newblock Immersion cooling of electronics in dod installations.
\newblock Technical Report LBNL-1005666, Berkeley National Laboratories, 2016.

\bibitem{jalilicost}
Majid Jalili, Ioannis Manousakis, {\'I}nigo Goiri, Pulkit~A Misra, Ashish
  Raniwala, Husam Alissa, Bharath Ramakrishnan, Phillip Tuma, Christian Belady,
  Marcus Fontoura, et~al.
\newblock Cost-efficient overclocking in immersion-cooled datacenters.
\newblock In {\em Proceedings of the International Symposium on Computer
  Architecture (ISCA'21)}, 2021.

\bibitem{Villa2020}
Herb Villa.
\newblock Liquid cooling vs. immersion cooling deployment.
\newblock
  \url{https://blog.rittal.us/liquid-cooling-vs-immersion-cooling-deployment},
  2020.

\bibitem{alibaba2018}
Alibaba.
\newblock Immersion cooling for green computing.
\newblock
  \url{https://www.opencompute.org/files/Immersion-Cooling-for-Green-Computing-V1.0.pdf},
  2018.

\bibitem{Varma2019}
Dhruv Varma.
\newblock Two-phase versus single-phase immersion cooling.
\newblock
  \url{https://www.grcooling.com/wp-content/uploads/2020/03/grc-blog-library-tech-comparison-\%E2\%80\%94-two-vs-single-phase-immersion-cooling.pdf},
  2019.

\bibitem{ramdas2019impact}
Shrinath Ramdas, Pavan Rajmane, Tushar Chauhan, Abel Misrak, and Dereje
  Agonafer.
\newblock Impact of immersion cooling on thermo-mechanical properties of pcb's
  and reliability of electronic packages.
\newblock In {\em International Electronic Packaging Technical Conference and
  Exhibition}, volume 59322. American Society of Mechanical Engineers, 2019.

\bibitem{microsoft2021}
Sabina Weston.
\newblock Microsoft is submerging servers in boiling liquid to prevent teams
  outages.
\newblock
  \url{https://www.itpro.co.uk/server-storage/datacentr/359129/microsoft-submerges-servers-in-boiling-liquid-toprevent-teams?amp},
  2021.

\bibitem{feelheat20}
{Voices of the Industry}.
\newblock Data centers feeling the heat! {T}he history and future of data
  center cooling.
\newblock
  \url{https://datacenterfrontier.com/history-future-data-center-cooling/},
  2020.

\bibitem{geng2014data}
Hwaiyu Geng.
\newblock {\em Data center handbook}.
\newblock John Wiley \& Sons, 2014.

\bibitem{shinde2019experimental}
Pravin~A Shinde, Pratik~V Bansode, Satyam Saini, Rajesh Kasukurthy, Tushar
  Chauhan, Jimil~M Shah, and Dereje Agonafer.
\newblock Experimental analysis for optimization of thermal performance of a
  server in single phase immersion cooling.
\newblock In {\em International Electronic Packaging Technical Conference and
  Exhibition}, volume 59322. American Society of Mechanical Engineers, 2019.

\bibitem{kanbur2020two}
Baris~Burak Kanbur, Chenlong Wu, Simiao Fan, Wei Tong, and Fei Duan.
\newblock Two-phase liquid-immersion data center cooling system: Experimental
  performance and thermoeconomic analysis.
\newblock {\em International Journal of Refrigeration}, 118:290--301, 2020.

\bibitem{DBLP:journals/tsusc/SonDCB17}
Jungmin Son, Amir~Vahid Dastjerdi, Rodrigo~N. Calheiros, and Rajkumar Buyya.
\newblock Sla-aware and energy-efficient dynamic overbooking in sdn-based cloud
  data centers.
\newblock {\em {IEEE} Trans. Sustain. Comput.}, 2(2):76--89, 2017.

\bibitem{haghshenas2020infrastructure}
Kawsar Haghshenas, Somayye Taheri, Maziar Goudarzi, and Siamak Mohammadi.
\newblock Infrastructure aware heterogeneous-workloads scheduling for data
  center energy cost minimization.
\newblock {\em IEEE Transactions on Cloud Computing}, 2020.

\bibitem{patankar2010airflow}
Suhas~V Patankar.
\newblock Airflow and cooling in a data center.
\newblock {\em Journal of Heat transfer}, 132(7), 2010.

\bibitem{tuma2010merits}
Phillip~E Tuma.
\newblock The merits of open bath immersion cooling of datacom equipment.
\newblock In {\em 2010 26th Annual IEEE Semiconductor Thermal Measurement and
  Management Symposium (SEMI-THERM)}, pages 123--131. IEEE, 2010.

\bibitem{ebrahimi2014review}
Khosrow Ebrahimi, Gerard~F Jones, and Amy~S Fleischer.
\newblock A review of data center cooling technology, operating conditions and
  the corresponding low-grade waste heat recovery opportunities.
\newblock {\em Renewable and Sustainable Energy Reviews}, 31:622--638, 2014.

\bibitem{li2015current}
Zhen Li and Satish~G Kandlikar.
\newblock Current status and future trends in data-center cooling technologies.
\newblock {\em Heat Transfer Engineering}, 36(6):523--538, 2015.

\bibitem{shah2016effects}
Jimil~M Shah, Richard Eiland, Ashwin Siddarth, and Dereje Agonafer.
\newblock Effects of mineral oil immersion cooling on it equipment reliability
  and reliability enhancements to data center operations.
\newblock In {\em 2016 15th IEEE Intersociety Conference on Thermal and
  Thermomechanical Phenomena in Electronic Systems (ITherm)}, pages 316--325.
  IEEE, 2016.

\bibitem{WATSON20172711}
Bryony Watson and Vinod~Kumar Venkiteswaran.
\newblock Universal cooling of data centres: A cfd analysis.
\newblock {\em Energy Procedia}, 142:2711--2720, 2017.
\newblock Proceedings of the 9th International Conference on Applied Energy.

\bibitem{nadjahi2018review}
Chayan Nadjahi, Hasna Louahlia, and St{\'e}phane Lemasson.
\newblock A review of thermal management and innovative cooling strategies for
  data center.
\newblock {\em Sustainable Computing: Informatics and Systems}, 19:14--28,
  2018.

\bibitem{kuncoro2019immersion}
IW~Kuncoro, NA~Pambudi, MK~Biddinika, I~Widiastuti, M~Hijriawan, and KM~Wibowo.
\newblock Immersion cooling as the next technology for data center cooling: A
  review.
\newblock In {\em Journal of Physics: Conference Series}, volume 1402, page
  044057. IOP Publishing, 2019.

\bibitem{alibaba2019}
Yangfan Zhong.
\newblock A large scale deployment experience using immersion cooling in
  datacenter.
\newblock Alibaba Group: Open Compute Project Summit, 2019.

\bibitem{Asperitas2021}
Asperitas.
\newblock Immersion cooling solutions for datacentres.
\newblock \url{https://www.asperitas.com/}, 2021.

\bibitem{ngoko2018future}
Yanik Ngoko, Nicolas Saintherant, Christophe Cerin, and Denis Trystram.
\newblock How future buildings could redefine distributed computing.
\newblock In {\em 2018 IEEE International Parallel and Distributed Processing
  Symposium Workshops (IPDPSW)}, pages 1232--1240. IEEE, 2018.

\bibitem{submer2021}
Submer.
\newblock Datacenters that make sense.
\newblock \url{https://submer.com/}, 2021.

\bibitem{Miller2014}
Rich Miller.
\newblock Inside {SuperNAP 8:} switch's {Tier IV} data fortress.
\newblock
  \url{https://www.datacenterknowledge.com/archives/2014/02/11/inside-supernap-8-switchs-tier-iv-data-fortress/},
  2014.

\bibitem{McNevin2013}
Ambrose McNevin.
\newblock Apac data center survey reveals high pue figures across the region.
\newblock \url{https://www.datacenterdynamics.com/}, 2013.

\bibitem{reddy2017metrics}
V~Dinesh Reddy, Brian Setz, G~Subrahmanya~VRK Rao, GR~Gangadharan, and Marco
  Aiello.
\newblock Metrics for sustainable data centers.
\newblock {\em IEEE Transactions on Sustainable Computing}, 2(3):290--303,
  2017.

\bibitem{jones2018information}
Nicola Jones.
\newblock The information factories.
\newblock {\em Nature}, 561(7722):163--6, 2018.

\bibitem{Trader2019}
Tiffany Trader.
\newblock {AMD Launches Epyc Rome}, {First 7nm CPU}.
\newblock \url{https://www.hpcwire.com/2019/08/08/amd-launches-epyc-rome-first-
  7nm-cpu/}, 2019.

\bibitem{gill2018taxonomy}
Sukhpal~Singh Gill and Rajkumar Buyya.
\newblock A taxonomy and future directions for sustainable cloud computing: 360
  degree view.
\newblock {\em ACM Computing Surveys (CSUR)}, 51(5):1--33, 2018.

\bibitem{dayarathna2015data}
Miyuru Dayarathna, Yonggang Wen, and Rui Fan.
\newblock Data center energy consumption modeling: A survey.
\newblock {\em IEEE Communications Surveys \& Tutorials}, 18(1):732--794, 2015.

\bibitem{shehabi2016united}
Arman Shehabi, Sarah Smith, Dale Sartor, Richard Brown, Magnus Herrlin,
  Jonathan Koomey, Eric Masanet, Nathaniel Horner, In{\^e}s Azevedo, and
  William Lintner.
\newblock United states data center energy usage report.
\newblock 2016.

\bibitem{gess2014investigation}
Joshua Gess, Sushil Bhavnani, Bharath Ramakrishnan, R~Wayne Johnson, Daniel
  Harris, Roy Knight, Michael Hamilton, and Charles Ellis.
\newblock Investigation and characterization of a high performance, small form
  factor, modular liquid immersion cooled server model.
\newblock In {\em 2014 Semiconductor Thermal Measurement and Management
  Symposium (SEMI-THERM)}, pages 8--16. IEEE, 2014.

\bibitem{tuma2010open}
Phil Tuma.
\newblock Open bath immersion cooling in data centers: A new twist on an old
  idea.
\newblock {\em Electronics Cooling}, page~10, 2010.

\bibitem{Neudorfer2016}
J.~{Neudorfer}, M.~{Ellsworth}, D.~{Kulkarni}, and H.~{Zien}.
\newblock Liquid cooling technology update.
\newblock {\em The Green Grid White Paper 70}, 2016.

\bibitem{Day2019}
T.~{Day}, P.~{Lin}, and R.~{Bunger}.
\newblock Liquid cooling technologies for data centers and edge applications.
\newblock {\em Schneider Electric White Paper 265}, 2019.

\end{thebibliography}




%

\newpage\mbox{} \newpage

\begin{IEEEbiography}[{\includegraphics[width=1in,height=1.25in,clip,keepaspectratio]{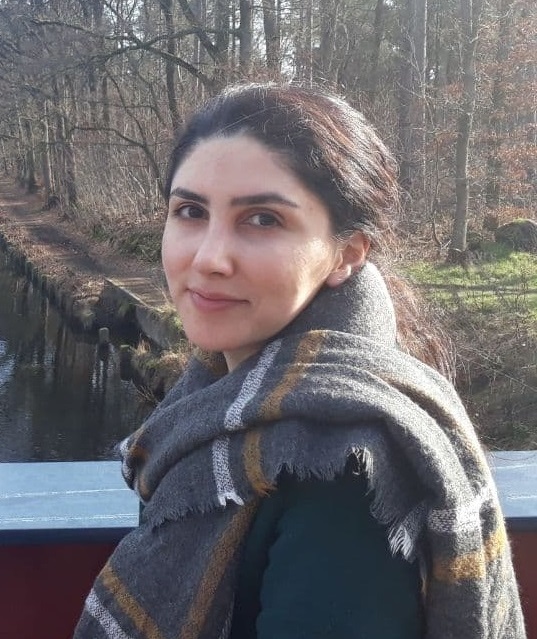}}]{Kawsar Haghshenas} is a post-doctoral researcher in the Smart Energy Systems (SES) laboratory at the University of Stuttgart. She received her Ph.D. degree in Computer Science from University of Tehran in 2020, M.Sc. degree in Computer Engineering from Sharif University of Technology in 2014, and B.Sc. degree in Computer Engineering from K.N.Toosi University of Technology in 2012. She was visiting the Embedded Systems Laboratory (ESL) at EPFL University from 2017 to 2018. Her research interests include system level energy optimization,  machine learning and workload scheduling in the area of data centers and cloud computing.
\end{IEEEbiography}

\begin{IEEEbiography}[{\includegraphics[width=1in,height=1.25in,clip,keepaspectratio]{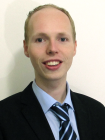}}]{Brian Setz} is a Research Assistant and PhD candidate at the Institute of Architecture of Application Systems (IAAS) of the University of Stuttgart, where he is a member of the Service Computing Department. He holds a masters degree in Computing Science from the University of Groningen. His research interests include: Internet of Things, Green Computing, Cloud Computing, and Energy Efficient Data Centers.
\end{IEEEbiography}

\begin{IEEEbiography}[{\includegraphics[width=1in,height=1.25in,clip,keepaspectratio]{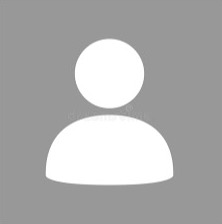}}]{Yannis Bloch} studied his B.Sc. degree in Computer Science at the Institute of Architecture of Application Systems (IAAS) of the University of Stuttgart.
\end{IEEEbiography}

\begin{IEEEbiography}[{\includegraphics[width=1in,height=1.25in,clip,keepaspectratio]{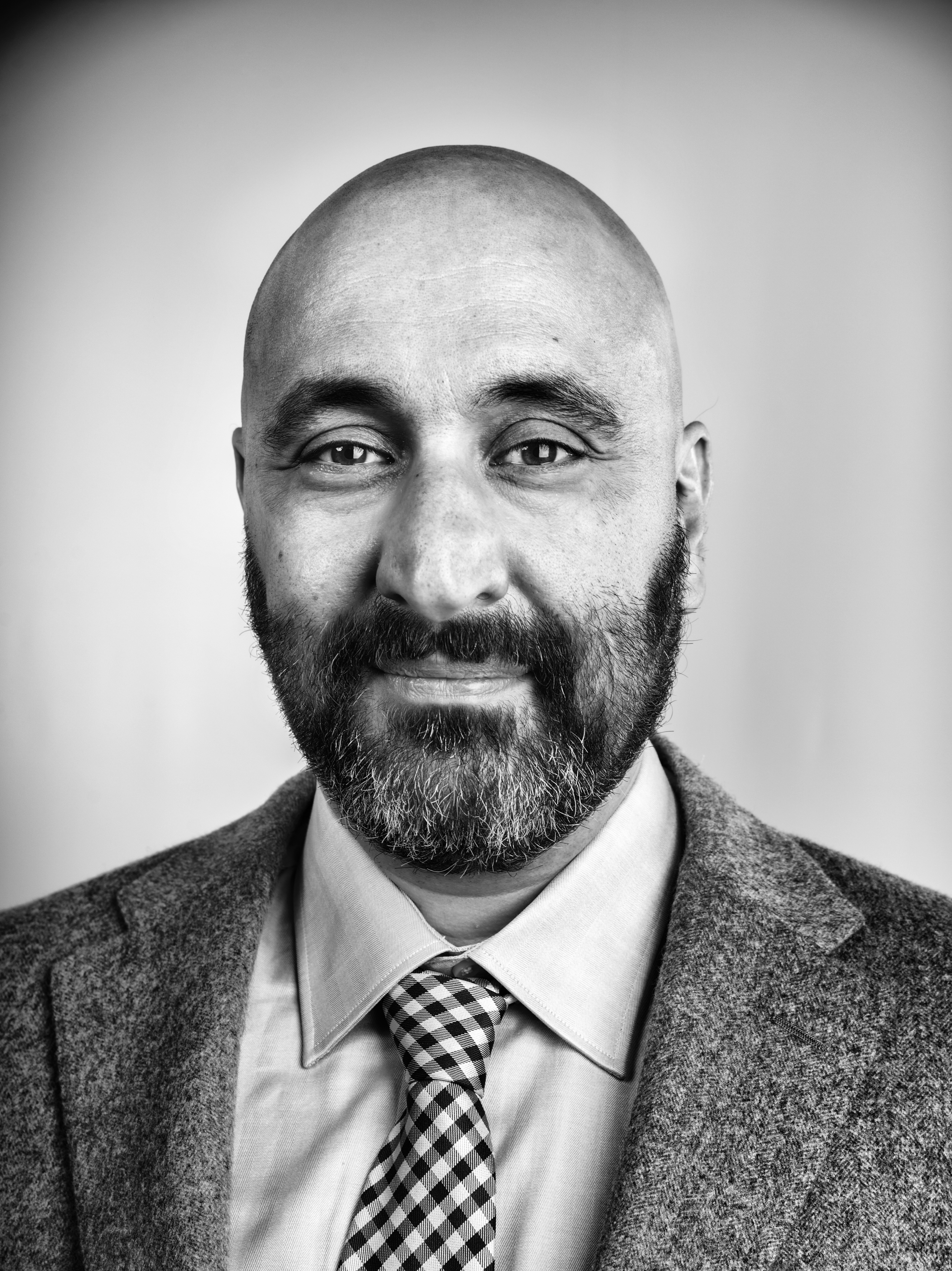}}]{Marco Aiello Shell}
Marco Aiello (SIEEE) is professor of Computer Science and Head of the Service Computing Department at the University of Stuttgart, Germany. He is also an elected member of the European Academy of Sciences and Arts and honorary professor of Distributed Systems at the University of Groningen, The Netherlands, where he was a faculty member from 2006 till 2018. He holds a PhD in Logic from the University of Amsterdam, the Habilitation in Applied Informatics from TU Wien, and a master degree in Engineering from La Sapienza University of Rome. His research interests are in Service Computing, Smart Energy Systems, and Data Centers. 
\end{IEEEbiography}

\end{document}